\documentclass[twocolumn,aip,apl,10pt]{revtex4-1}

\usepackage{graphicx}
\usepackage{dcolumn}
\usepackage{bm}
\usepackage{amsmath,amssymb}

\def\U#1{{%
\def\O{\mbox{O}}
\def\u{\mbox{u}}
\mathcode`\u=\mu
\mathcode`\O=\Omega
\mathrm{#1}}}
\def\degree{\mbox{$^\circ$}}
\def\ii{{\mathrm{i}}}
\def\vct#1{{\mathchoice{\mbox{\boldmath$#1$}}{\mbox{\boldmath$#1$}}%
  {\mbox{\scriptsize\boldmath$#1$}}{\mbox{\scriptsize\boldmath$#1$}}}}
\def\sub#1{_{\scriptsize\mbox{#1}}}

\begin{document}

\title{A linear-to-circular polarization converter with
   half transmission and half reflection using a single-layered 
   metamaterial}

\author{Yasuhiro Tamayama}
\email{tamayama@vos.nagaokaut.ac.jp}
\author{Kanji Yasui}
\affiliation{Department of Electrical Engineering, Nagaoka University of
   Technology, Nagaoka 940-2188, Japan}
\author{Toshihiro Nakanishi}
\author{Masao Kitano}
\affiliation{Department of Electronic Science and Engineering, Kyoto
   University, Kyoto 615-8510, Japan}

\date{\today}

\begin{abstract}

A linear-to-circular
polarization converter with half transmission and half reflection 
using a single-layered metamaterial is theoretically and numerically demonstrated.
The unit cell of the metamaterial consists of two coupled split-ring
resonators with identical dimensions. A theoretical
analysis based on an electrical circuit model of the coupled
split-ring resonators indicates that the linear-to-circular
polarization converter is achieved when the magnetic coupling between the
split-ring resonators is set to a certain strength. 
A finite-difference time-domain simulation reveals
that the single-layered metamaterial 
behaves as the linear-to-circular polarization converter and that 
the polarization converter has the combined characteristics of 
a half mirror and a quarter-wave plate.

\end{abstract}



\maketitle

Polarization is one of the important characteristics of electromagnetic
waves. 
The manipulation of polarization is essential for
optical communication, sensing, and other applications.\cite{saleh07} 
Fundamental optical elements such as 
waveplates, polarizers, and polarization rotators 
are used to manipulate polarization.
These elements are fabricated mainly 
using anisotropic media, chiral media,
the Brewster effect, and the Faraday effect.\cite{saleh07}

Metamaterials have been demonstrated to enable the development of useful polarization
control devices that cannot be achieved using naturally occurring media. 
The electromagnetic response of metamaterials can be controlled by 
design of the
shape, material, and configuration of the unit structure. In addition,
metamaterials are geometrically scalable; therefore, 
exotic polarization control devices can be fabricated for electromagnetic waves
from radio to optical frequencies using metamaterials.
For example, cross polarization
converters,\cite{hao07,li10_apl,grady13,cong13,cheng13_apl,ginzburg13} 
linear-to-circular polarization converters,\cite{labadie10,yu12,ginzburg13,jiang13,wang14} 
achromatic wave plates,\cite{nagai14}
and broadband circular polarizers\cite{gansel09}
have been realized over a broad frequency range. 
Metamaterials with giant optical
activity,\cite{gonokami05,rogacheva06,wang_b09_apl,zhang09,liu07_prb,li08_apl,cheng13_apa}
which can be used as 
thin polarization rotators, have also been studied.

In this Letter, we theoretically and numerically 
demonstrate that a linear-to-circular polarization converter
with half transmission and half reflection can be realized using a
single-layered metamaterial composed of coupled resonators. 
Metamaterials composed of coupled resonators have been intensively
investigated for the control of electromagnetic 
waves.\cite{liu07_prb,li08_apl,cheng13_apa,nakanishi12,czaplicki13,zhang_prl08,tassin_prl09,liu_nat09,tamayama12,nakanishi13,tamayama14}
A metamaterial with characteristics similar to that employed in this work has been 
previously reported.\cite{jiang13}
However, no theory for the structural design of such a metamaterial 
has been reported. 
The theory is necessary for the further development of 
electromagnetic wave control using metamaterials.
An electrical
circuit model of the metamaterial unit structure is used to 
theoretically show how the linear-to-circular polarization converter 
with half transmission and half reflection can be realized. 
The characteristics of the designed metamaterial based on the theory
are then numerically analyzed to confirm that the metamaterial 
does behave as the linear-to-circular
polarization converter. 
The polarization converter consists of a single-layered metamaterial and 
has combined characteristics of
a half mirror and a quarter-wave plate. 
This study is expected to lead to the development of single-layered metamaterials that
have combined characteristics of multiple 
fundamental optical elements, 
which could thus contribute to the miniaturization of optical systems.

\begin{figure}[tb]
\begin{center}
\includegraphics[scale=0.65]{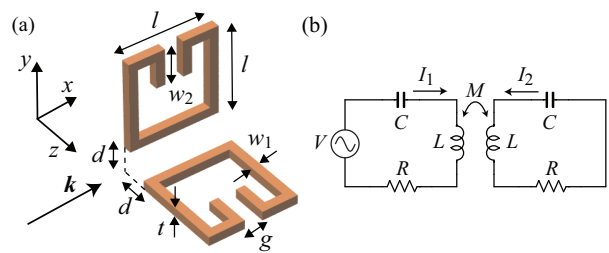}
\caption{(a) Schematic of the unit structure of the metamaterial and (b)
 electrical circuit model of the unit structure. }
\label{fig:structure}
\end{center}
\end{figure}

\begin{figure*}[tb]
\begin{center}
\includegraphics[scale=0.85]{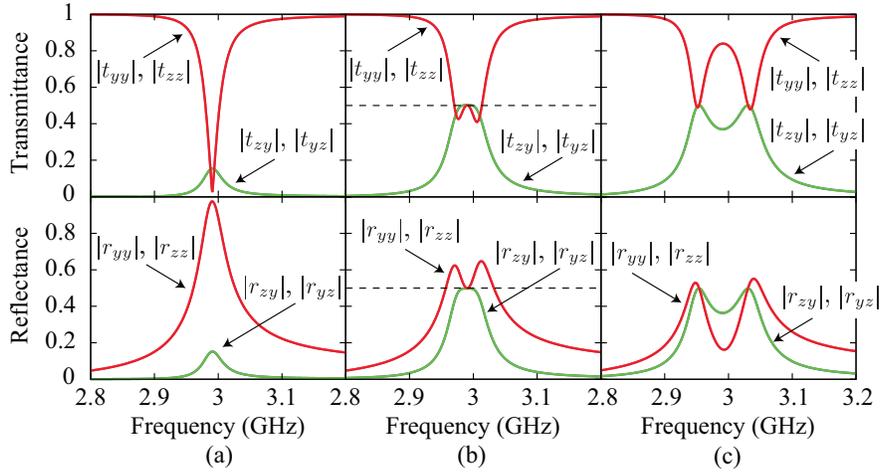}
\caption{Transmission spectra (upper panel) and reflection spectra (lower
 panel) for (a) $d=10\,\U{mm}$, (b) $8\,\U{mm}$, and (c)
 $6\,\U{mm}$. The horizontal dashed lines in (b) represent 
 transmittance and reflectance of $0.5$.}
\label{fig:3spectra}
\end{center}
\end{figure*}

A linear-to-circular polarization converter with half transmission and
half reflection can be realized using a single-layered
metamaterial with the unit structure 
shown in Fig.\,\ref{fig:structure}(a). The unit structure
of the metamaterial is composed of two identical split-ring resonators. 
The axes of the rings are orthogonal to each other and to the direction
of the wavevector of the incident electromagnetic wave. 
The unit structure is periodically arranged in the $y$ and $z$
directions. 

First, the theory of the linear-to-circular polarization
converter is described using the electrical circuit model of the metamaterial unit
structure shown in Fig.\,\ref{fig:structure}(b). 
The electrical circuit model consists of two identical 
inductor-capacitor series
resonant circuits coupled with each other via a mutual inductance. Each
inductor-capacitor series resonant circuit corresponds to the split-ring
resonator, and the mutual inductance represents the magnetic coupling
between the two split-ring resonators. The voltage source represents the
$y$ or $z$ polarized incident electromagnetic wave. 

Applying Kirchhoff's voltage law to the electrical circuit yields 
$I_1 = -ZV/[Z^2 +(\omega M)^2]$ and 
$I_2 = -\ii \omega M V / [Z^2 + (\omega M)^2]$, 
where $\omega$ is the angular frequency of the voltage source and
$Z=R-\ii \{ \omega L - [1/(\omega C)] \}$. 
For $\omega = \omega_0 = 1/\sqrt{LC}$ and $|M| = R / \omega_0$, 
we obtain $I_1 = -V/(2R)$ and $I_2 = -\ii (M/|M|) [V/(2R)]$, which is
reduced to 
$I_2 / I_1 = \ii M / |M|$. This implies that 
the radiations from the two split-ring resonators in each unit cell have
the same amplitudes and a phase difference of $\pi /2$. 
That is, the metamaterial radiates a circularly polarized electromagnetic
wave when the $y$ or $z$ polarized
electromagnetic wave with $\omega = \omega_0$ is incident on 
the metamaterial with $|M| = R / \omega_0$. 
This phenomenon yields the linear-to-circular polarization
converter with half transmission and half reflection, as described in the next paragraph. 

\begin{figure*}[tb]
\begin{center}
\includegraphics[scale=0.85]{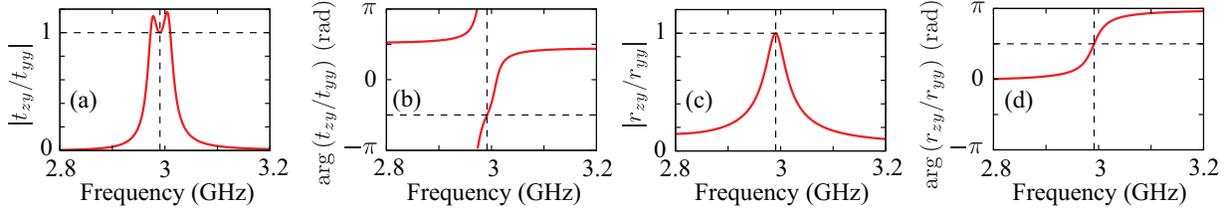}
\caption{Frequency dependence of (a) $|t_{zy}/t_{yy}|$, (b)
 $\arg{(t_{zy}/t_{yy})}$, (c) $|r_{zy}/r_{yy}|$, and (d)
 $\arg{(r_{zy}/r_{yy})}$ for $d=8\,\U{mm}$. The vertical dashed line in
 each figure represents $f_0 = 2.99\,\U{GHz}$.}
\label{fig:ratio}
\end{center}
\end{figure*}

\begin{figure*}[tb]
\begin{center}
\includegraphics[scale=0.85]{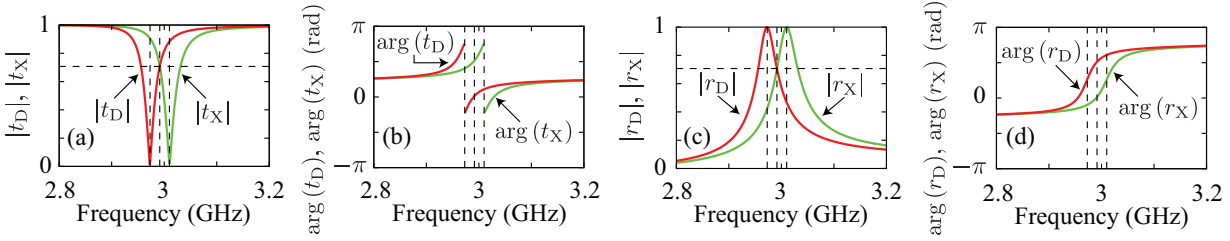}
\caption{Frequency dependence of (a) the absolute values and 
(b) arguments
 of the transmittances for the D and X polarizations 
and (c) and (d) those of the reflectances for the D and X polarizations
for $d=8\,\U{mm}$. The three vertical dashed lines in each figure 
represent, from left to right, $f_1 = 2.97\,\U{GHz}$, 
$f_0 = 2.99\,\U{GHz}$, and $f_2 = 3.01\,\U{GHz}$. 
The horizontal dashed lines represent $1/\sqrt{2}$.}
\label{fig:45deg}
\end{center}
\end{figure*}

We consider the transmittance and reflectance of the metamaterial for
$\omega = \omega_0$ and $|M| = R/\omega_0$. The co-polarized (cross-polarized) 
transmittance and reflectance are defined 
as $t_1$ and $r_1$ ($t_2$ and $r_2$), respectively. 
The reflected wave
consists of only the radiation from the metamaterial; thus, 
$r_2 / r_1 = I_2 / I_1 = \ii M / |M|$. 
That is, the reflected wave is 
a circularly polarized wave. On the other hand, the transmitted wave
is a superposition of the incident wave and the radiation from the
metamaterial. It is necessary for the calculation of the transmittance
to evaluate the relation between the 
amplitude of the incident wave and that of 
the wave radiated from the metamaterial. To evaluate
the amplitude, the radiation from the metamaterial is considered for
$M=0$. In the case of $M=0$, currents $I_1$ and $I_2$
are reduced to $I_1 = -V/R = I_{10}$ 
and $I_2 =0$ for $\omega = \omega_0$. When an
electromagnetic wave with $\omega = \omega_0$ is incident on a resonant
metamaterial (Lorentz medium) with a sufficiently strong response, 
the resonant metamaterial radiates an electromagnetic wave, which results in the disappearance of the transmitted wave. 
This implies that the radiated wave caused by current
$I_1 = I_{10}$ has the same amplitude and opposite phase to the incident
wave. For $\omega = \omega_0$ and $|M|=R/\omega_0$, 
the transmittances and reflectances satisfy 
$t_1 =1/2$, $|t_2| = |r_1| = |r_2| = 1/2$, and 
$t_2 / t_1 = I_2/(-I_1) = -\ii M/|M|$,
because $I_1 = I_{10} /2$, $I_2 / I_1 = \ii M / |M|$, and 
$\arg{(t_1)} =-\arg{(I_1)}$. 
Therefore, both of the transmitted and reflected waves are circularly
polarized waves with half the power of the incident wave. 
The transmitted and reflected
circularly polarized waves have the same helicity, because 
$t_2/t_1 = -r_2/r_1 = -\ii M/|M|$ 
and the propagation directions of these two waves
are opposite.

The transmission and reflection characteristics of the
metamaterial shown in Fig.\,\ref{fig:structure}(a) were analyzed using a 
finite-difference time-domain (FDTD) method\cite{taflove05} to confirm
the theory based on the electrical circuit model. 
The geometrical parameters of the metamaterial shown in
Fig.\,\ref{fig:structure}(a) were $l = 12\,\U{mm}$, 
$w_1 = 1\,\U{mm}$, $w_2 = 4\,\U{mm}$, $g = 2\,\U{mm}$, 
and $t = 1\,\U{mm}$. 
The periods of the unit cell in the $y$ and $z$ directions
were set to $48\,\U{mm}$. The separation
$d$ between the two split-ring resonators was set to $10\,\U{mm}$,
$8\,\U{mm}$,
or $6\,\U{mm}$. The FDTD simulation space was discretized into 
uniform cubes with dimensions of 
$1\,\U{mm} \times 1\,\U{mm} \times 1\,\U{mm}$.

Figure \ref{fig:3spectra} shows the transmission and reflection spectra 
for $d=10\,\U{mm}$, $8\,\U{mm}$, and $6\,\U{mm}$. 
Here, $t_{ij}$ ($r_{ij}$) ($i$, $j = y$, $z$) represents 
the complex amplitude of the $i$ polarized component of the 
transmitted (reflected) wave when the $j$ polarized wave with unity
amplitude is incident on the metamaterial.
The calculated results satisfy
$t_{yy} = t_{zz}$, 
$t_{zy} = t_{yz}$, $r_{yy} = r_{zz}$, and $r_{yz} = r_{zy}$;
thus, we focus on the transmittances and reflectances for 
the $y$ polarization incidence until the end of the next paragraph. 
The splitting of the resonance dip in the transmission
spectrum (resonance peak in the reflection spectrum) decreases with
an increase in $d$. For $d=10\,\U{mm}$, the splitting disappears, and the
transmission and reflection spectra are almost the same as those for
$M=0$ (not shown). 
For $d=8\,\U{mm}$, $| t_{zy} | \simeq | t_{yy} | \simeq 0.5$ and 
$| r_{zy} | \simeq | r_{yy} | \simeq 0.5$ are satisfied at around
$f= f_0 = 2.99\,\U{GHz}$, which is the resonance frequency of the
split-ring resonator. That is, half the power of the incident wave is
transmitted and the other half is 
reflected. The absolute values of the transmittances and
reflectances satisfy the necessary conditions for the linear-to-circular
polarization converter. 

We now examine whether the linear-to-circular
polarization conversion with half transmission and half reflection is
achieved at $f=f_0$ for $d=8\,\U{mm}$. 
Figure \ref{fig:ratio} shows the frequency dependence of the absolute
values and arguments of $t_{zy}/t_{yy}$ and $r_{zy}/r_{yy}$
for $d=8\,\U{mm}$. 
The absolute value of $t_{zy}/t_{yy}$ takes a local minimum value 
of $1.00$ and that of $r_{zy}/r_{yy}$ takes a maximum value of $0.997$ at 
$f=f_0$. The arguments of $t_{zy}/t_{yy}$ and $r_{zy}/r_{yy}$ are
 equal to $-0.500\pi$ and $0.500\pi$, respectively, at $f = f_0$. 
This implies that almost ideal
circularly polarized waves are transmitted and reflected 
when the $y$ or $z$ polarized wave is incident
on the metamaterial. 
These results confirm that 
the linear-to-circular polarization
converter with half transmission and half reflection can be achieved
using the metamaterial designed based on the electrical circuit model. 

As we described above, $t_{yy} = t_{zz}$, 
$t_{zy} = t_{yz}$, $r_{yy} = r_{zz}$, and $r_{yz} = r_{zy}$ are
satisfied for the metamaterial, which 
implies that the eigenpolarizations in the metamaterial
are the $\pm 45\degree$ polarizations. This observation is confirmed from
the Maxwell equation. The wavevector $\vct{k}$ is in the $x$
direction; thus, only $\varepsilon_{lm}$ and $\mu_{lm}$ ($l$, $m=y$, $z$)
in the permittivity and permeability tensor components of the
metamaterial are necessary for the analysis. 
Taking the symmetry of the metamaterial into account, the components of
the permittivity and permeability tensors are, respectively, written as 
$\varepsilon_{lm} = \varepsilon \delta_{lm}$ and 
$\mu_{lm} = \mu \delta_{lm} + \mu\sub{c} (1-\delta_{lm})$, where 
$\delta_{lm}$ is the Kronecker delta.
After substitution of these tensors
into the Maxwell equation and diagonalization, 
we obtain 
$
k 
[
E_{\mathrm{D}+} ~ 
E_{\mathrm{X}+} ~
E_{\mathrm{D}-} ~
E_{\mathrm{X}-}
]^T
=
( \omega \varepsilon )
\mathrm{diag}
(
Z\sub{D},\,
Z\sub{X},\,
- Z\sub{D},\,
- Z\sub{X}
)
[
E_{\mathrm{D}+} ~
E_{\mathrm{X}+} ~
E_{\mathrm{D}-} ~
E_{\mathrm{X}-}
]^T$,
where $E_{\mathrm{D}\pm} = E_y + E_z \mp Z\sub{D} ( H_y - H_z ) $, 
$E_{\mathrm{X}\pm} = E_y - E_z \pm Z\sub{X} ( H_y + H_z ) $, 
$Z\sub{D} = \sqrt{(\mu - \mu\sub{c})/\varepsilon}$,
$Z\sub{X} = \sqrt{(\mu + \mu\sub{c})/\varepsilon}$, and
$T$ denotes matrix transposition. 
This equation indicates that the eigenpolarizations in the
metamaterial are the $45\degree$ polarization (D polarization) and
the $-45\degree$ polarization (X polarization), and that
the difference between the wavenumbers for the $\pm 45\degree$ polarizations
is caused by the magnetic coupling $\mu\sub{c}$ between
the split-ring resonators. 

Figure \ref{fig:45deg} shows the transmission and reflection spectra for
the $\pm 45 \degree$ polarized waves for $d=8\,\U{mm}$. 
First, the physical meaning of 
the resonant response of the metamaterial is
discussed based on the spectra.  
The splitting of the resonance line in
Fig.\,\ref{fig:3spectra} is
derived from resonances at $f=f_1 = 2.97\,\U{GHz}$ for the D
polarization and $f=f_2 = 3.01\,\U{GHz}$ for the
X polarization in Fig.\,\ref{fig:45deg}. 
The direction of the magnetic field is in the
$-45\degree$ direction ($45\degree$ direction) for the D polarization (X 
polarization). Thus, the resonance mode at $f=f_1$ ($f=f_2$) is the symmetric
(antisymmetric) mode, the current flow of which in the coupled
split-ring resonator is symmetric (antisymmetric) with respect to the
symmetry plane of the unit structure. For the vertical or horizontal
polarization incidence, both of the two kinds of the resonance modes can
be excited. Therefore, when $f_2-f_1$ is much larger (smaller) than the
resonance linewidth of the split-ring resonator, which is proportional
to $R$, the two kinds of the resonance modes are observed as distinct
resonance lines (one resonance line) in the transmission and reflection
spectra for the vertical or horizontal
incidence.  
Next, the characteristics of the metamaterial at $f=f_0$ are described on
the basis of the $\pm 45 \degree$ polarizations.  
The absolute values of transmittances and reflectances for 
the $\pm 45 \degree$ polarized waves are $1/\sqrt{2}$ at 
$f = f_0$. That is, the power transmittances and 
reflectances for $\pm 45 \degree$ polarized waves are $1/2$ at
$f = f_0$. 
The difference between the arguments of the transmittances (reflectances)
for the $\pm 45 \degree$ polarizations are $\pi /2$ at $f=f_0$. 
These characteristics imply that the metamaterial behaves
as a half mirror with quarter-wave retardation at $f=f_0$. 
Therefore, when the $y$ or $z$ polarized wave, of which the polarization axis
is $45\degree$ different from the fast and slow axes (D and X axes)
of the quarter-wave plate,
is incident on the metamaterial, circularly polarized waves with half
the power of the incident wave are transmitted and reflected. 
If the coupling strength between the two split-ring resonators becomes
stronger (weaker), the difference between $f_1$ and $f_2$ increases
(decreases). In this case, the phase difference between the
eigenpolarizations is varied from $\pi /2$, so that the transmitted
and reflected waves become elliptically polarized waves. In addition, the
amplitude of the transmitted wave becomes different from that of the
reflected wave. 
This observation is consistent with the theory based on
the electrical circuit model that 
describes the phenomenon on the basis of the $0\degree$ and
$90\degree$ polarizations.

We have demonstrated that a linear-to-circular polarization converter with half
transmission and half reflection can be realized using 
a single-layered metamaterial
composed of coupled split-ring resonators. A theoretical analysis
based on the electrical circuit model revealed that the 
linear-to-circular polarization converter is achieved for $\omega = \omega_0$ and 
$|M| = R / \omega_0 $. An FDTD simulation demonstrated that 
$|M| = R / \omega_0 $ can be realized for coupled split-ring resonators
and that circularly polarized electromagnetic waves with the same power
and helicity are transmitted and reflected for the
vertical or horizontal polarization incidence. 
The eigenpolarizations in
the metamaterial were found to be the $\pm 45\degree$ polarizations. 
The difference between the transmittance (reflectance) arguments for 
the eigenpolarizations are $\pi / 2$, and the power transmittances and
power reflectances for the eigenpolarizations are $1/2$ at the resonance
frequency of the split-ring resonators, which implies that the
metamaterial has the combined characteristics of a half mirror and a
quarter-wave plate. 
If meta-atoms with a low quality factor and sufficiently strong response
are used as elements for the present metamaterial, then the frequency
dependence of the metamaterial characteristics at around 
$\omega = \omega_0$
becomes small, and the linear-to-circular polarization converter with
small deviations of the transmittance and reflectance from the ideal
can be achieved for a relatively broad
frequency range. 
The result of the present study suggests that compact
circular polarization
beam splitters\cite{jacobs88,davis01,azzam05,tamayama08} 
and half beam splitters may be fabricated using
metamaterials composed of coupled resonators. 

This research was supported by a Grant-in-Aid for Scientific Research on
Innovative Areas (No.\@ 22109004) from the Ministry of Education,
Culture, Sports, Science, and Technology of Japan, and by a Grant-in-Aid
for Research Activity Start-up (No.\@ 25889028) from the Japan Society
for the Promotion of Science. 

%

\end{document}